# Martingale Option Pricing

J.L. McCauley[+], G.H. Gunaratne[++], and K.E. Bassler

Physics Department
University of Houston
Houston, Tx. 77204
jmccauley@uh.edu

[+]Senior Fellow
**COBERA**
Department of Economics
*J.E.Cairnes Graduate School of Business and Public Policy*
NUI Galway, Ireland

[++]Institute of Fundamental Studies
Kandy, Sri Lanka

### Abstract

We show that our earlier generalization of the Black-Scholes partial differential equation (pde) for variable diffusion coefficients is equivalent to a Martingale in the risk neutral discounted stock price. Previously, the equivalence of Black-Scholes to a Martingale was proven for the case of the Gaussian returns model by Harrison and Kreps, but we prove it for much a much larger class of returns models where the returns diffusion coefficient depends irreducibly on both returns x and time t. That option prices blow up if fat tails in logarithmic returns x are included in market return is also proven.

## 1. Introduction

In order to motivate the consideration of diffusive models, we begin with a summary of the known facts about financial markets discovered in our recent data analysis based on six years of Euro/Dollar intraday FX prices at 1 min. time intervals [1]: (i) Finance markets are nonstationary with no observable approach to statistical equilibrium, ruling out asymptotically stationary processes like those considered in [2], e.g. (ii) To the extent that reliable histograms can be constructed, the market distribution $f(x,t)$ is exponential in x where x is the log return. (iii) Increments in $x(t)$ are nonstationary, which is characteristic of drift free (or negligible drift) Markov processes whenever the variance is not linear in t [3]. (iv) The Hurst exponent is $H \neq 1/2$, but scaling is not observed over the entire range of data and H is not even the same within the four separate intraday scaling regions. (v) The variance is finite, ruling out Levy distributions. (vi) Autocorrelations in increments vanish rapidly for times t>10 min., ruling out fractional Brownian motion and pointing again to Markov dynamics. These facts suggest stochastic dynamics with an (x,t) diffusion coefficient, where $x = \ln p/p_c$ is the log return, p is the price and $p_c$ is a reference price [4]. We therefore consider Markovian markets [5], markets consisting entirely of nonGaussian noise. The form of the noise that describes the market is determined by the diffusion coefficient $D(x,t)$, a function that must be deduced from real market data. The diffusion coefficient reflects the behavior of the noise traders [4], who provide the liquidity in normal, heavily traded markets [6].

This paper discusses fair option prices for nontrivial diffusive models, models where the diffusion coefficient depends inseparably on both x and t. Easy examples are linear diffusion $D(x,t)=t^{2H-1}(1+x/t^H)$, leading to exponential processes, and quadratic diffusion $D(x,t)= t^{2H-1}(1+\varepsilon x^2/t^{2H})$, leading to student-t like distributions [4,5]. *Our main result is that the generalization of the Black-Scholes partial differential equation (pde) to the case of variable diffusion D(x,t) describes a Martingale in the risk neutral discounted stock price.* This was proven by Harrison and Kreps [7] for the original Black-Scholes model [8], where D=constant (see also [9,10]) so that returns are Gaussian. Our interest is in classes of nonGaussian returns models that reflect the empirical data faithfully. Extending the original Black-Scholes result to a purely time dependent diffusion $D(t)$ is trivial, the returns in that case are still Gaussian. Steele [9] discusses Black-Scholes pdes with Martingale solutions for the case where $D(x)$ depends on x alone, but that restriction eliminates the market.

Martingale solutions of diffusive pdes with D(x,t)≠constant are discussed by instructively Durrett [11]. We present for the first time the proof that the Green function for the Black-Scholes type pde with diffusion coefficient D(x,t) depending on both x and t describes a Martingale. The Green function provides, therefore, the so-called 'Martingale measure' of interest in financial engineering. That this is so is not widely understood, if at all.

For Markovian markets the stock price p(t) is described by a stochastic differential equation (sde) [5], as is the returns variable x(t). The Fokker-Planck pde, whose solution is the market Green function [12], transforms one to one with the returns sde. We've shown [12] that the Black-Scholes pde has the same Green function as the market Fokker-Planck pde, but with the stock interest rate replaced by the bank interest rate. Financial engineers and mathematicians have not realized that the market Green function is the Martingale measure that they prefer to construct from the standpoint of sdes via Girsanov's theorem. The application of Girsanov's theorem to sdes is nontrivial when variable diffusion coefficients D(x,t) are considered [11]. In any case, Girsanov is not needed, we get the desired Martingale directly from the Black-Scholes pde.

Many papers [13] and even some books [14] have claimed to price options with fat tailed distributions. Here, we will prove rigorously that the option price diverges if the required Green function has fat tails. In particular, fat tails in any market distribution will appear in the Green function, contrary to the (unproven) folklore of financial engineering.

We begin by reviewing the connection between the Black-Scholes pde and the Fokker-Planck pde describing the market Green function, a connection that we discovered and reported some years ago [12,15].

## 2. The Black-Scholes PDE and Kolmogorov's First PDE

A delta hedge strategy based on a nontrivial diffusion coefficient D(x,t) is locally risk neutral [12], just as in the case of the original Black-Scholes-Merton model [8] where D=constant (Gaussian returns). To describe variable diffusion models we begin with the sde for the stock price p(t) at present time t,

$$dp = \mu p dt + \sqrt{p^2 d(p,t)} dB(t), \quad (1)$$

where B(t) is the Wiener process and µ is the (hard to estimate) stock 'interest rate'. The quantity $p^2 d(p,t)$ is the price diffusion coefficient. The function d(p,t) is constant in the Black-Scholes model (lognormal pricing, or Gaussian returns) but real market data [1] forces us to consider models where d(p,t)≠constant. A merely t-dependent function d(t), independent of p, is trivially lognormal as well, as one can easily see by transforming to logarithmic returns.

We next transform the sde (1) to returns $x = \ln p(t)/p_c$ where $p_c$ is a reference price. Using Ito's lemma we obtain

$$dx = (\mu - D(x,t)/2)dt + \sqrt{D(x,t)} dB, \quad (2)$$

where D(x,t)=d(p,t). The solution of the corresponding Fokker-Planck pde

$$\frac{\partial g(x,t;x_o,t_o)}{\partial t} = -\frac{\partial}{\partial x}((\mu - D(x,t)/2)g(x,t;x_o,t_o)) + \frac{1}{2}\frac{\partial^2}{\partial x^2}(D(x,t)g(x,t;x_o,t_o))$$
(3)

with g(x,t;x',t)=δ(x-x') is the *market Green function*, the transition probability density (or conditional 2-point density) for the Markov process described by (2,3). The empirical market density (the 1-point density) is then given by f(x,t)=g(x,t;0,0), a quantity that one hopes to extract as histograms from financial time series [1].

Consider next the delta hedge, a portfolio long on Δ shares of stock and short a call w. If the portfolio is chosen to increase in value at the bank (money market, CD, …) interest rate then Δ=w', where prime denotes differential with respect to p, then we have the risk neutral portfolio where the Green function for the Black-Scholes pde [12,15]

$$rw = \frac{\partial w}{\partial t} + rp\frac{\partial w}{\partial p} + \frac{d(p,t)p^2}{2}\frac{\partial^2 w}{\partial p^2}, \quad (4)$$

is to be used to solve the forward time initial value problem for the option price. The derivation of (4) from the delta hedge condition is exactly the same as in the original case [8] where d(p,t)=constant.

Consider a European call with strike price K and expiration date T. Transforming $w=e^{r(t-T)}g_p$ we get

$$0 = \frac{\partial g_p}{\partial t} + rp\frac{\partial g_p}{\partial p} + \frac{d(p,t)p^2}{2}\frac{\partial^2 g_p}{\partial p^2}. \quad (5)$$

The fair price of the call is then predicted for a normal, liquid market to be given by the initial value solution

$$C(p, K, T, t) = e^{r(t-T)} \int_{-\infty}^{\infty} dp_T(p_T - K)\theta(p_T - K)g_p(p_T, T; p, t), \quad (6)$$

where p is the known stock price at present time t, $p_T$ is the unknown price at expiration, and $g_p(p_T,T;p,T)=d(p_T-p)$. The Green function transforms like a probability density, so with the transformation u(x,t)dx=w(p,t)dp to log returns x=lnp/$p_c$, we get the time-transformed generalized Black-Scholes pde in the form [12]

$$0 = \frac{\partial u}{\partial t} + (r - D(x,t)/2))\frac{\partial u}{\partial x} + \frac{D(x,t)}{2}\frac{\partial^2 u}{\partial x^2}. \quad (7)$$

This is a very beautiful result: *This pde is exactly the backward time equation, or first Kolmogorov equation, corresponding to the Fokker-Planck pde (the second Kolmogorov equation) (3) for the market Green function g if we choose µ=r in the latter* [11]. With the choice µ=r then both pdes are solved by the same Green function g, so that no information is provided by solving the Black-Scholes pde (4) that is not already contained in the Green function of the Market F-P equation (3). To be explicit, according to the theory of backward time integration [16,17] we must understand the time-transformed Black-Scholes pde (7) as

$$0 = \frac{\partial u(x_o,t_o)}{\partial t_o} + (r - D(x_o,t_o)/2)\frac{\partial u(x_o,t_o)}{\partial x_o} + \frac{D(x_o,t_o)}{2}\frac{\partial^2 u(x_o,t_o)}{\partial x_o^2}$$
(8)

where $u(x_o,t_o) = g(x,t;x_o,t_o)$ solves the Fokker-Planck pde (3) in (x,t). This is a result that physicists should love: everything of interest can be calculated once one has the Green function.

Using $g_p dp = g dx$ because the Green function transforms like a (conditional) probability density, we can now use the returns Green function g to price calls (and puts as well) risk neutrally as

$$C(p,K,T-t) = e^{r(t-T)} \int_{-\infty}^{\infty} (p_T - K)\theta(p_T - K) g(x_T,T;x,t) dx_T, \quad (9)$$

where $x_T = \ln p_T/p_c$ is based on the a priori unknown price $p_T$ at expiration, and $x = \ln p/p_c$ where p is the price at present time t. Unlike the original Black-Scholes model, the stock interest rate does not disappear entirely from the option price: it is hidden irreducibly in the time dependence of the consensus price $p_c$, which locates the peak of the distribution and is the dynamic generalization to reality of the neo-classical definition of 'value' in a hypothetical stationary market [4]. In practice [12,15], one can impose a condition that fixes μ in the consensus price by the bank interest rate r.

The empirical density of returns is given by $g(x,t;0,0) = f(x,t)$, i.e., this is the correct 1-point density if we start with a return $x_o = 0$ at $t_o = 0$. If in the call price (9) we restrict to x=0, so that $p \approx p_c$ and then approximate the drift $R(x,t) = \mu - D(x,t)/2 \approx$ constant in (3) and use the exponential density with H=1/2, then we obtain the formulae that we used much earlier [12,15] to price options empirically correctly. This means that we've approximated an arbitrary stock price p at present time t<T by the consensus price $p_c$, a result would be expected to be of practical use only so long as stock prices don't vary too much. first, correct option pricing is not at all a reliable test of the correctness of an underlying model of market dynamics. Second, anyone who has watched Euro/Dollar FX rates on a quasi-daily basis over the last seven years (like one of the authors, who lives on Dollars eight months/year in Austria) knows that the Dollar fluctuations within a day are typically on the order of a few cents on a scale of roughly one Euro.

## 3. Martingale Option Pricing

We show next that the generalized B-S pde (4) is equivalent to a Martingale in the risk neutral discounted stock price. The B-S pde is equivalent via a time transformation to the backward time Kolmogorov pde

$$0 = \frac{\partial u}{\partial t} + (r - D/2)\frac{\partial u}{\partial x} + \frac{D}{2}\frac{\partial^2 u}{\partial x^2} \quad . \quad (10)$$

The call price is calculated from the Green function $u=g^{+}(x,t;x_T,T)$ of this pde, where the dagger denotes the adjoint. The forward time Kolmogorov pde

$$\frac{\partial g}{\partial T} = -\frac{\partial}{\partial x_T}((r - D(x_T,T)/2)g) + \frac{\partial^2}{\partial x_T^2}(\frac{D(x_T,T)}{2}g) \quad (11)$$

has exactly the same Green function $g(x_T,T;x,t)=g^{+}(x,t;x_T,T)$. The price sde corresponding to this Fokker-Planck pde (dropping subscripts capital T, for convenience) is

$$dp = rpdt + \sqrt{p^2 d(p,t)}dB \quad (12)$$

where $d(p,t)=D(x,t)$ and r is the risk neutral rate of return. With $y=x-rt$ and $g(x,x';t,t')=G(y,y';t,t')$ (since $dx=dy$) we obtain

$$\frac{\partial G}{\partial t} = -\frac{\partial}{\partial y}(-\frac{E}{2}G) + \frac{\partial^2}{\partial y^2}(\frac{E}{2}G) \quad (13)$$

with $E(y,t)=D(x,t)$, which has the sde

$$dy = -E(y,t)dt/2 + \sqrt{E(y,t)}dB(t) \quad (14)$$

and yields the corresponding price sde (with $x=\ln S(t)/S_c$)

$$dS = \sqrt{S^2 e(S,t)} dB(t) \qquad (15)$$

with price diffusion coefficient $e(S,t)=E(y,t)=D(x,t)=d(p,t)$. This shows that the risk neutral discounted price $S=pe^{-rt}$ is a Martingale.

Girsanov's theorem is often stated in financial math texts [9,18] as transforming a Wiener process plus a drift term into another Wiener process. This is wrong: when the drift depends on a random variable x and is not merely t-dependent, then the resulting process is not Wiener. Durrett [11] proves Girsanov's theorem in the generality needed for empirically based option pricing: one starts with any Martingale x(t) (e.g., dx=√D(x,t)dB generates a Martingale), adds an (x,t) dependent drift term R(x,t), and then constructs a new Martingale. The new Martingale is *not* a Wiener process unless (a) the old process is Wiener, and (b) the drift is not a function of a random variable x(t). The sole exception to this rule is the lognormal sde $dx=\mu x dt+\sigma_1^2 x dB$ with $\sigma_1$ constant, which is trivially 'Wiener' by the simple coordinate transformation y=lnx.

That the Black-Scholes pde is equivalent to a Martingale in the risk neutral discounted stock price was proven abstractly and nontransparently for the case of the Gaussian returns model [7]. Our proof above is transparent and is not restricted to the unphysical assumption that D(x,t) is independent of x.

**4. Option Pricing with Fat Tails**

Consider the price of a call for x>δ where u=(x-δ)/√t, where δ is determined by the consensus price $p_c$ [4,19]:

$$C(p,K,T-t) = e^{r(t-T)} \int_{\ln K/p}^{\infty} (p_T - K) g(x_T, T; x, t) dx_T, \qquad (16)$$

and p is the known stock price at present time t<T. We know the Green function both empirically and analytically only for the case where g(x,t;0,0)=f(x,t). This yields

$$C(p_c, K, T-t) = e^{r(t-T)} \int_{\ln K/p_c}^{\infty} (p_T - K) f(x_T, T) dx_T, \qquad (17)$$

and this special case is adequate for making our next point: *if* the observed density (the empirical density) has fat tails $f(x,t) \approx x^{-\mu}$ for $x \gg 1$, so we get

$$C(p_c, K, T-t) \approx e^{r(t-T)} \int_{\ln K/p_c}^{\infty} pe^x x^{-\mu} dx = \infty! \quad (18)$$

This result is exact. If one inserts a finite cutoff then the option price is very sensitive to the cutoff, meaning that one can then predict essentially *any* option price.

Many papers can be found in the literature and on the web purporting to price options using fat tails. All are in serious error in one way or another. In Borland [20], there are several mathematical mistakes in the analysis. In that model the diffusion coefficient is [5]

$$D(x,t) = (c(2-q)(3-q))^{2H-1} t^{2H-1}(1+(q-1)x^2/C^2(q)t^{2H}) = t^{2H-1}\hat{D}(u) \quad (19)$$

where $u = x/t^H$, $H = 1/(3-q)$, and

$$C(q) = c^{(q-1)/2(3-q)}((2-q)(3-q))^H \quad (20)$$

where

$$c^{1/2} = \int_{-\infty}^{\infty} du(1+(q-1)u^2)^{1/(1-q)} . \quad (21)$$

One mistake is that in the attempt to construct a price Martingale using this quadratic coefficient a term $\int x^2(t)/t^{2H} dt$ appears in the exponential factor derived from an attempted application of Girsanov's Theorem. That term was treated incorrectly as $\int x^2(t)/t^{2H} dt \propto x^2(T)/T^{2H}$, where T is the expiration time for the option and the average is over the return x(T) at expiration. That mistake gave rise to a spurious Gaussian convergence factor that does not appear in the correct option pricing formula.

That the substitution is wrong is easy to see: equations (71) and (143) in ref. [20] assert that

$$x(t)/t^H = x(T)/T^H \quad (22)$$

where (translating the different notations) Borland's variable $\Omega$ is our variable x. In our notation, this is equivalent to asserting that u(t)=u(T). By Ito's lemma the sde for the random variable u(t) is

$$du = (-Hu/t + t^{-H}(\mu - D(x,t)/2))dt + t^{-H}\sqrt{D(x,t)}dB(t) \quad (23)$$

or

$$du = (-Hu/t + t^{-H}(\mu - t^{2H-1}\widehat{D}(u)/2))dt + t^{-1/2}\sqrt{\widehat{D}(u)}dB(t). \quad (24)$$

The stochastic integral is given by

$$u(T) = u(t) + \int_t^T (-Hu(s)/s + s^{-H}(\mu - s^{2H-1}\widehat{D}(u(s))/2))ds + \int_t^T s^{-1/2}\sqrt{\widehat{D}(u(s))}dB(s) \quad (25)$$

so that u(T)≠u(t).

Summarizing the two new results of this paper, first, we've proven that the Green function for the Black-Scholes pde generalized to (x,t) dependent diffusion coefficients D(x,t) describes a Martingale in the risk neutral discounted stock price. Second, we've show rigorously that option prices blow up at the infinite upper limit if fat tails are included, and have pointed out that fat tails in the 'Martingale measure' will reflect fat tails in the market returns distribution, in contrast with misplaced expectations in financial engineering. The amount of money in the world is finite, if hard to determine (due to credit), but the predicted option price will then be very sensitive to the choice of cutoff.

**Acknowledgement**

KEB is supported by the NSF through grants #DMR-0406323 and #DMR-0427938, by SI International and the AFRL, and by TcSUH. GHG is supported by the NSF through grant #PHY-0201001 and by TcSUH. JMC is grateful to Harry Thomas for pointing out the Fokker-Planck pde for the variable u, and to Lisa Borland for stimulating email correspondence. We're also grateful to a referee for several references and for encouraging us to write the introduction.